# Characterization of $Nb_{22}O_{54}$ microrods grown from niobium oxide powders recovered from mine tailings

B. Sotillo[a,*], F. A. López[b], L. Alcaraz[b], P. Fernández[a]

[a] Department of Materials Physics, Faculty of Physics, Complutense University of Madrid, 28040, Madrid, Spain.

[b] Centro Nacional de Investigaciones Metalúrgicas (CENIM-CSIC), Avda. Gregorio del Amo, 8, 28040 Madrid, Spain

[*] Corresponding author: bsotillo@ucm.es; +34 913944121

**Abstract.** In this work, the possibility of using niobium oxide recovered from tailings from the Penouta Sn-Nb-Ta deposit (located at the North of Spain) as starting material for growing microstructures is demonstrated. The properties of the starting material have been studied to understand its crystal structure, quality and purity. Recovered niobium oxide powders are mainly of $TT$-$Nb_2O_5$. These powders have been used to grow $Nb_{22}O_{54}$ microrods by an evaporation method in an argon atmosphere. Different characterization techniques (X-Ray diffraction, scanning electron microscopy, micro-Raman spectroscopy, luminescence) have been used to determine the properties of $Nb_{22}O_{54}$ microrods, mainly focusing on the crystal quality and refractive index. The present study opens the way to the transformation of waste (mine tailings) into a material of high technological value as niobium oxide, and its reintroduction into the value chain for a wide range of applications, from coatings to batteries and supercapacitors.

### Introduction

Niobium oxides constitute a group of materials that has proved their versatility for different applications. Among them, the high dielectric constant of some of the phases make them useful in the fabrication of capacitors [1], competing with tantalum oxide. Recently it has been shown that orthorhombic niobium pentoxide has a good value of pseudocapacitance, which makes it a good candidate to be used in the fabrication of supercapacitors and batteries [2-7]. In particular, non-stoichiometric phases of niobium oxide ($Nb_{22}O_{54}$, $Nb_{12}O_{29}$) have been less reported in the literature, but they have been proposed as good candidates for intercalation electrode materials in lithium-ion batteries [8-10]. Along with these applications that have attracted high interest, niobium oxides are also useful for photocatalysis [11, 12], electrocatalysis [13-15], coatings or light guiding [16], due to a bandgap in the near UV range or a high refractive index, respectively. It has to be taken into account that many of the properties that make niobium oxide interesting for applications are highly dependent on the crystal structure of the material as well as on the stoichiometry [16], so deeper studies are needed to elucidate the physical phenomena behind this dependence.

Another problem faced by the development of niobium-oxide based devices is the reduced amount of niobium found in the Earth, obtained from the Coltan mineral coming from regions that are not politically stable. Many strategic materials, like Niobium, Tantalum and Rare Earth Elements (REE) are included in the last list of materials considered critical by the European Union. The last list available dates from 2017 and include 27 critical raw materials due to their importance for high-tech products and emerging innovations and the risk in the security of supply and economic importance [17]. The extraction of these metals is also expensive in terms of energy: up to 30% of the energy required to refine these metals is used in separation processes, like liquid-liquid extraction [18, 19]. Along with the scarcity or criticality of many of the materials used in our technological society, humans must promote the recovery of the huge amount of waste. In this sense, one of the EU's priorities is to promote the transition to a circular economy, where the materials and products manufactured with them are kept in the life cycle as long as possible.

In this paper it is shown the possibility of obtaining good quality niobium oxide from the tailings of mining deposits, located in the north of Spain, that were initially focused on the extraction of Sn [20]. It is possible to recover this precious material (niobium oxide) even if the amount of material in the mine is small. And following the philosophy of the circular economy, the recovered material have been used to grow high quality $Nb_{22}O_{54}$ microrods with good properties for the different applications described above.

**Methods**

Penouta mine tailings (cassiterite and columbotantalite) are treated by a pyrometallurgical process to obtain a metal tin ingot and a slag[20]. The slag was treated by acid leaching, using $HF/H_2SO_4$ as leaching agent. The extraction of Nb in leaching aqueous phase were performed using 35% (v/v) Cyanex 923® diluted in Solvesso with high yields 97.7% for Nb. The stripping solutions used for the selective stripping and separation of niobium had compositions $NH_4F$ 0.3 M and $NH_3$ 0.1 M respectively. After the separation of Nb, solid precursors were obtained by precipitation. For the synthesis of niobium compounds, ammonia, $[NH_3]$=17.7 M was employed. After precipitation, solids were filtered, washed with deionized water and dried at 80ºC for 12 hours. Nb precursor obtained by precipitation with $NH_3$ it was mainly an amorphous compound, corresponding with a hydrated niobium oxide ($Nb_2O_5 \cdot nH_2O$). Nb precursor was calcined in a tubular furnace at 1200°C during 4 hours in $N_2$ atmosphere, to obtain niobium oxide. Figure 1 shows schematically the steps of the synthesis process. A deeper description of this process and the properties of the recovered material can be found in [21].

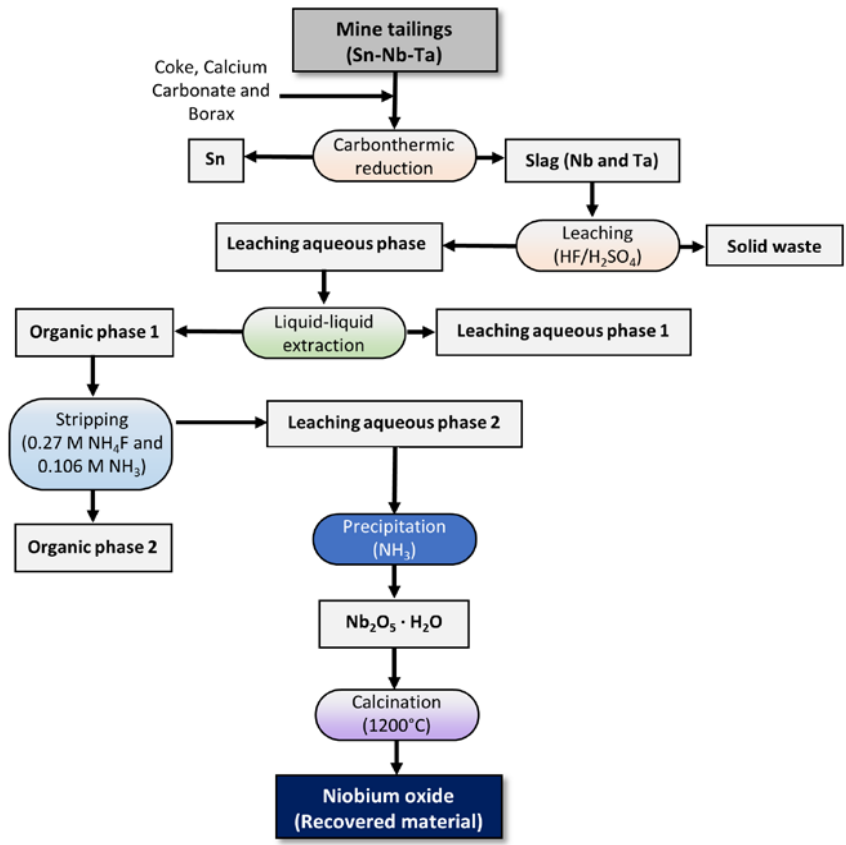

**Figure 1.** Scheme of the synthesis of niobium oxide [21]

The goal of this work is the use of the niobium oxide recovered from the mine tailings to grow microrods. The recovered material is obtained in the form of powders. This starting powders were compacted under a compressive load to form disk-shaped pellets of about 8 mm in diameter and 1 mm in thickness. The pellets were then placed on an alumina boat and introduced in a muffle oven under a continuous Ar flow. Gas flow is varied between 1 and 2.5 l/min, and treatment duration between 1 and 20 hours. Two different treatment temperatures have been selected: 1200°C and 1300°C. In all cases the temperature was raised to the final value at a rate of 10°C/min.

The microrods obtained have been characterized using several characterization techniques, already described in[22]. First, X-Ray Diffraction measurements (XRD) have been done by means of a PANalytical Empyrean diffractometer using Cu-K$\alpha$ radiation, with a step in 2$\theta$ of 0.05°. Then, Scanning Electron Microscopy (SEM) based techniques have been applied: secondary electron mode, x-ray microanalysis (EDX) and electron backscattered diffraction (EBSD). A FEI Inspect or a LEICA 440 SEM have been used for emissive mode measurements. For performing the spectroscopic measurements on the microrods, they have been separated from the pellets and deposited on a silicon <100> substrate. For EDX measurements a Bruker AXS Quantax attached to the LEICA SEM has been used and EBSD measurements have been carried out with a Bruker e⁻Flash Detector coupled to the FEI Inspect SEM working at 20 kV. The analysis of the EBSD data has been performed with ESPRIT QUANTAX CrystAlign commercial software. Micro-Raman measurements were carried out in a confocal microscope Horiba JobinYvon LABRAM-HR, using the 632.8 nm line of a He-Ne laser. The laser was focused onto the sample with a 100× Olympus objective (0.9 NA), and the scattered light was also collected using the same objective (backscattering configuration). The grating used to analyse the signal had 1800 l/mm, and the signal was collected with an air-cooled charge coupled device camera (CCD). µRaman spectra were collected and analysed using the Labspec 5.0 software of the confocal microscope. Photoluminescence (PL) measurements were performed with 325 nm light from a He-Cd laser, using the same setup as for the Raman measurements, just changing the objective (a Thorlabs LMU-40×-NUV objective (0.5 NA)) and the grating (600 l/mm). Cathodoluminescence (CL) measurements were done Hitachi S2500 SEM. For collecting the CL emission, a HAMAMATSU PMA-12 CCD (measurement range between 200 and 900 nm), coupled to an optical fibre, has been used. All the measurements were done at room temperature.

The objective of using all these characterization techniques is to perform a deep study of the starting material and the microrods, in order to understand the changes produced as a consequence of the thermal treatments and to clarify their potential applications.

**Results and discussion**

X-Ray Diffraction (XRD) measurements have been performed on the recovered material to elucidate the crystal phase of the powders (Figure 2(a)). The obtained XRD pattern is shown in Figure 2(b). The most intense reflections can be associated with the pseudohexagonal phase of $Nb_2O_5$ (TT-$Nb_2O_5$, here it is used the nomenclature first proposed by Schäfer et. al in [23]) with $a=b=$ 3.6070 Å and $c =$ 3.9250 Å (ICDD no. 00-028-0317, black marks in Figure 2). There are also less intense reflections that can be ascribed to a non-stoichiometric monoclinic phase, specifically, $Nb_{12}O_{29}$ with space group A2/m and lattice parameters $a =$ 15.6856 Å; $b =$ 3.8307 Å; $c =$ 20.7100 Å and β= 113.06º (ICDD no. 04-014-6587, red marks in Figure 2). Two of the recorded peaks, located at 20.50° and 25.85°, may be related to crystalline $SiO_2$ with hexagonal crystal structure (ICDD no. 01-081-1665). The presence of these impurity is quite plausible as one of the components of the deposit is quartz. Further description of the recovered niobium oxide can be found in [21]. For the work presented here, TT-$Nb_2O_5$ phase is very interesting because other crystal phases can be obtained from it by applying thermal treatments [16]. TT-$Nb_2O_5$ is reported to crystallize in a "pseudohexagonal" or monoclinic crystal structure, but the lattice is not fully refined in the literature [16]. It can have different types of distorted niobia polyhedra, with six, seven or eight oxygen atoms coordinated to each Nb atom (tetragonal, pentagonal or hexagonal bypiramids), being some oxygen atoms missing to maintain the stoichiometry [24].

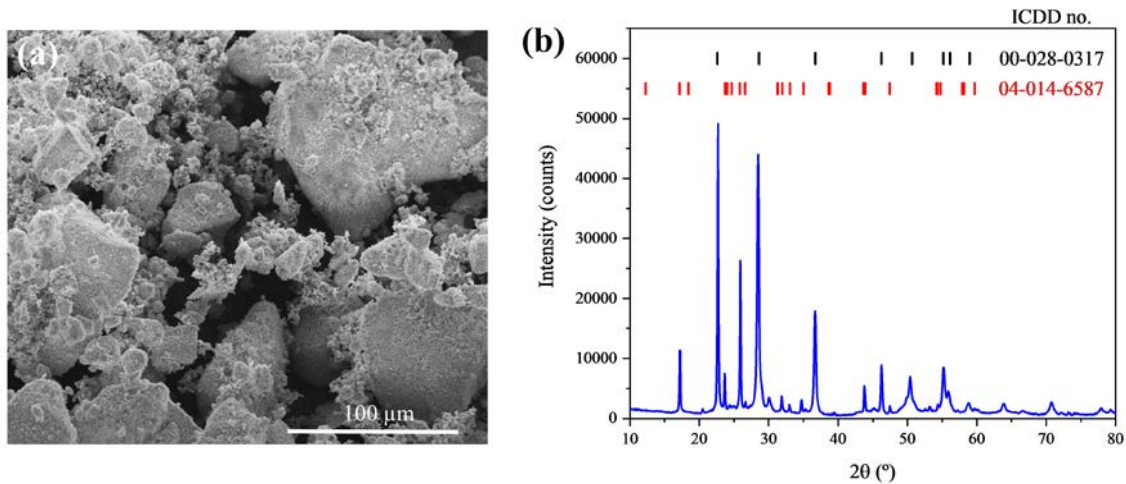

**Figure 2.** (a) SEM image of the niobium oxide powders recovered from mine tailing. (b) XRD pattern of the recovered material. The position of the peaks for the TT-$Nb_2O_5$ phase (black marks) and the monoclinic $Nb_{12}O_{29}$ phase (red marks) are indicated.

When pellets of the recovered material are treated at 1200ºC, a small density microrods start to grow on the surface of the pellets (Figure 3(a)). After 10 hours treatments, the rods, have typically lengths of tens of microns and widths of few microns, however there is no clear evidence of an increase in the length and/or density of these structures for longer treatments (up to 20 h). It is only when the temperature is increased to 1300ºC, when a high density of microrods can be obtained at the edges of the pellet even for 1 hour of treatment, as shown in Figure 3(b). These rods have a rectangular cross-section (Figure 3(c)), with side lengths in the range of a few microns. The number and length of these microrods vary depending on the duration of the thermal treatment. When the time is increased to 5 and 10 hours, all the pellet appear covered with this type of structures (Figure 3(d)). For the 10 h treatment, the length of the rods can be larger than 700 μm, reaching 1 mm in some cases. Increasing the time from 10 to 20 hours does not produce great differences in the number and length of the structures obtained. The section size of the rods do not show a significant change with the duration of the thermal treatment.

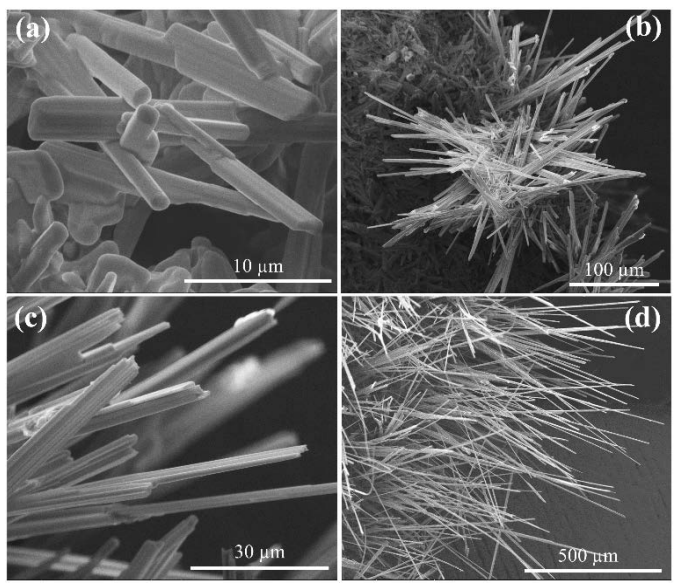

**Figure 3.** SEM images on niobium oxide microrods obtained at (a) 1200ºC and (b)-(d) 1300ºC.

XRD performed on the samples after the thermal treatments show that non-stoichiometric phases can be stabilized upon thermal treatment in Ar flux. At 1200ºC, the main phase detected is monoclinic $Nb_{12}O_{29}$ (Figure 4(a)), with the same crystallographic parameters indicated above. At 1300ºC, the phase detected $Nb_{22}O_{54}$ is also monoclinic, but belongs to a different space group, P2/m, and has lattice parameters $a$ = 15.7491 Å; $b$ = 3.8236 Å; $c$ = 17.8521 Å and β= 102.3º (ICDD no. 04-014-9203, purple marks in Figure 4(b)). For both set of samples (treated at 1200 or 1300ºC), some reflections can only be ascribed to the high temperature monoclinic phase of $Nb_2O_5$ (H-$Nb_2O_5$, ICDD no. 00-037-1468, green marks in Figure 4). When the temperature is raised to 1300ºC, the change from $Nb_{12}O_{29}$ to $Nb_{22}O_{54}$ is visible in the shift of the peak at 24.7º towards 24.9º (this peak cannot be associated to the H-$Nb_2O_5$ phase) [25], as shown in Figure 4(c).

The main building blocks for H-$Nb_2O_5$, $Nb_{12}O_{29}$ and $Nb_{22}O_{54}$ polymorphs (Figure 4 (d)-(f)) are octahedra formed by a niobium atom surrounded by six oxygen atoms ($NbO_6$). The crystal system is monoclinic, and the structures are then constructed by joining the octahedra in different manners, i.e. sharing corners to all surrounding octahedra or sharing edges. Typically, the sharing-edges octahedra will be more distorted than the sharing-corners octahedral [26]. Some $NbO_4$ tetrahedra can also appear to help to fill space (for example, at the corners of the cell in the H-$Nb_2O_5$ structure). The three structures differ in the oxygen content and in the crystal lattice parameters, and may be obtained performing high temperature treatments by varying the Nb:O ratio available in the treatment [25]. A complete description can be found in [23-25, 27].

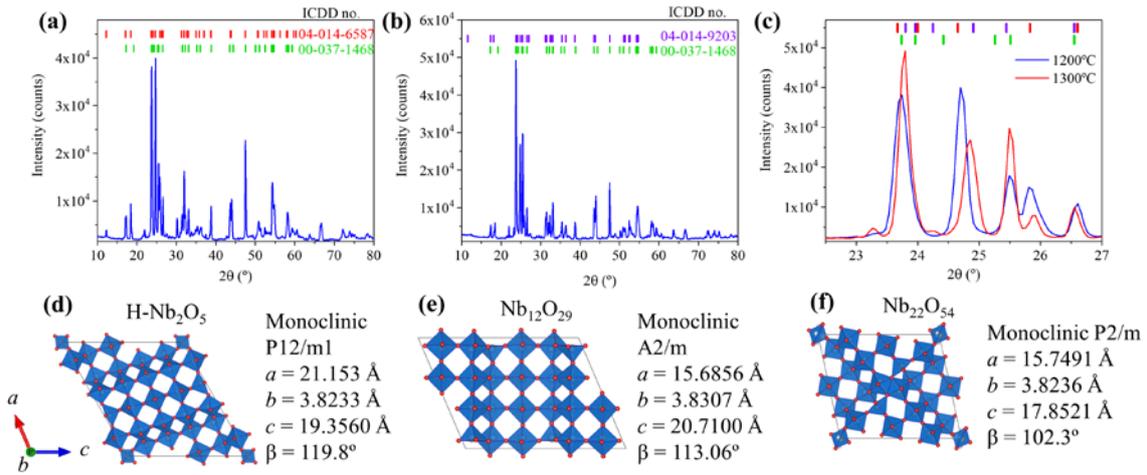

**Figure 4.** XRD pattern of the material treated at (a) 1200ºC and at (b) 1300ºC. (c) Comparison of XRD patterns for the two temperatures. The crystal phases identified are: monoclinic $Nb_{12}O_{29}$ (red marks, ICDD no. 04-014-6587); monoclinic $H-Nb_2O_5$ (green marks, ICDD no. 00-037-1468); and monoclinic $Nb_{22}O_{54}$ (purple marks, ICDD no. 04-014-9203). Structural schemes of the different crystallographic monoclinic phases of niobium oxide studied: (d) $H-Nb_2O_5$; (e) $Nb_{12}O_{29}$; (f) $Nb_{22}O_{54}$. The structures have been plotted using VESTA software [28]. Oxygen atoms are shown in red.

After performing the thermal treatment, there is no sign of incorporation (within the detection limit of EDX technique) in the microrods of other elements rather than Nb or O. This is an indication that the niobium oxide rods have a good purity. From XRD measurements, it is inferred the presence of a mixture of non-stoichiometric monoclinic phases and monoclinic $H-Nb_2O_5$. In order to confirm the crystal phase of the obtained structures, EBSD measurements have been performed on individual microrods (Figure 5). A Kikuchi pattern obtained from a rod is presented in Figure 5(a). It has been fitted by the CrystAlign software to the $Nb_{22}O_{54}$ monoclinic phase (the simulated pattern is shown below the measured one). By recording the Kikuchi pattern on a selected region of the rods, the crystal orientation can be determined. The inverse pole figure (IPF) maps for the three X, Y, Z directions are presented in Figure 5(b). These maps show, in a color scale, the main crystallographic directions that are closer to the X, Y, Z directions of the sample. From these results it can be seen that the growth direction of the rods is along the *b* axis, i.e. the short axis of the monoclinic structure (Figure 5(c)). The same growth direction has been observed in all the microrods tested. The other niobium oxide phases detected in XRD (Figure 4) are very likely related to the material remaining in the pellet.

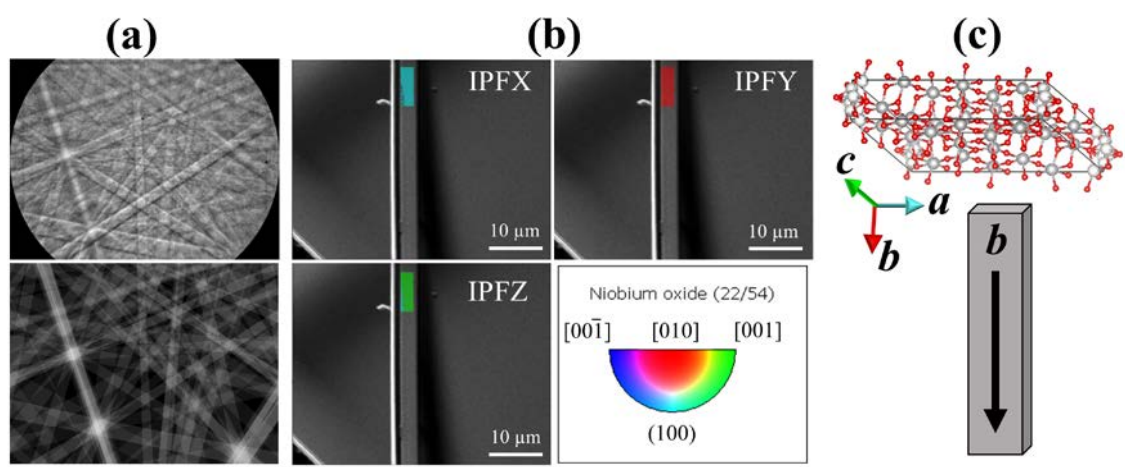

**Figure 5.** EBSD measurements performed on a niobium oxide microrods. (a) Measured (up) and simulated (down) Kikuchi pattern of a rod, associated to the monoclinic $Nb_{22}O_{54}$ phase. (b) IPF maps, superposed on SEM images of the rod, for the three axis X, Y, Z. (c) $Nb_{22}O_{54}$ crystal structure and its orientation in the obtained rods.

The change in the crystal phase from the starting material to the microrods is detected also in the micro-Raman spectra (Figure 6). The typical spectra recorded on the recovered niobium oxide material is labelled as "Recovered_1". It is composed of three main bands centred at 235, 300, and 693 cm$^{-1}$. These three bands can be associated with lattice vibrations of the TT-phase [29]. The two bands located in the low wavenumber region (235 and 300 cm$^{-1}$) are associated to the bending modes of the Nb-O-Nb linkages[29-31], whereas the band 690 cm$^{-1}$ is assigned to the symmetric stretching mode of niobia polyedra[29]. It is also possible, in the recovered material, to detect sharper peaks appearing at 130, 252, 634, 877 and 993 cm$^{-1}$ (spectrum labelled as "Recovered_2"). These sharper peaks are characteristic of monoclinic niobium oxide phases [24, 26, 29], also detected in the XRD pattern.

Peaks measured in the Raman spectra of the material treated at both 1200ºC and 1300ºC can be associated to vibrations of monoclinic phases, most of them related to internal vibrations of the niobia polyhedral [26]. The peaks are centred at 113, 132, 162, 264, 350, 480, 555, 634, 660, 827 and 993 cm$^{-1}$. The band at 993 cm$^{-1}$ is associated to the symmetric stretching modes of the terminal bonds of Nb=O in $NbO_6$ octahedra with a high degree of distortion [24, 29, 32]. The appearance of these type of bonds is related to the existence of shear planes [33], which appear in the H-$Nb_2O_5$ structure as well as in the $Nb_{22}O_{54}$ and $Nb_{12}O_{29}$ structures [25]. Corner-shared octahedral $NbO_6$ forms a Nb-O-Nb collinear bonding [24, 29] that produces the band located at 827 cm$^{-1}$. The bands that appear between 500 and 700 cm$^{-1}$ are related to vibration of slightly distorted octahedra $NbO_6$ structure [29]. The peaks ascribed to Nb-O-Nb angle deformations[24, 29] are found between 200 and 350 cm$^{-1}$, whereas between 350 and 500 cm$^{-1}$ the vibrations are related to the $ONb_3$ structure [24, 34]. Finally, the bands below 200 cm$^{-1}$ can be related to external lattice vibrations (for which the octahedra is considered as a rigid unit) [26]. The band at 113 cm$^{-1}$ could be related to a Nb-Nb mode [34]. The peak detected at 520 cm$^{-1}$ is associated to the silicon substrate.

As the $Nb_{22}O_{54}$ microrods are mostly monocrystalline, as confirmed by EBSD measurements, the Raman spectra obtained from the microrods are highly polarization sensitive, as can be seen by comparing the Raman spectrum of a rod with the growth axis parallel or perpendicular to the polarization of the incident laser (Figure 6). Modes that are more sensitive to this effect are those located at 993, 555, 264 and 113 cm$^{-1}$.

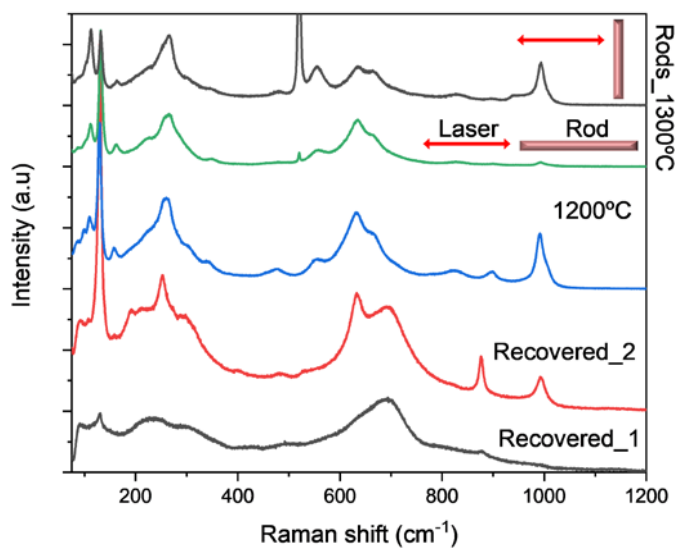

**Figure 6**. Micro-Raman spectra recorded on the recovered material (recovered_1 and recovered_2) on the material annealed at 1200ºC and on the rods obtained at 1300ºC. For the rods obtained at 1300ºC, the effect of the orientation of the rods respect to the laser polarization is shown.

Cathodoluminescence measurements have been performed on the recovered material as well as on obtained rods (Figure 7). In the recovered material, a broad band centred at about 405 nm is detected (Figure 7(a)). For the microrods, a similar band is also obtained, but the maxima are slightly blue shifted towards 390 nm (Figure 7(b)). The possible origin of this band is scarcely reported in the literature. The reported band gap of niobium pentoxide is typically located at higher energy (around or higher than 3.4 eV), and depends on the crystals structure and the stoichiometry of the oxide[35]. A strong blue emission located at similar wavelength has been associated to niobia octahedra complexes in ternary niobate materials [35-37], so it is possible that this emission is related to these complexes, present also in our material.

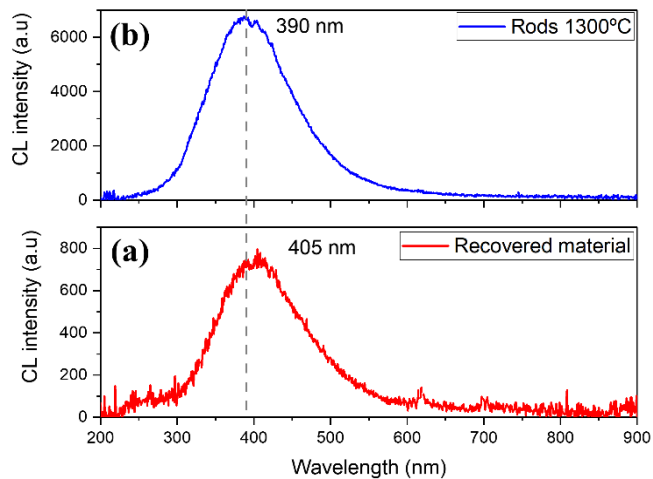

**Figure 7.** CL spectra recorded at room temperature on (a) recovered material; (b) microrods.

Finally, the refractive index has been determined by the light guiding and optical resonant cavity behavior of the obtained rods. (Figure 8). First, using a 632.8 nm laser, one of the sides of the structure is illuminated, and the guided light is recorded in an optical image using a 100× objective (Figure 8(a)). It is evident that the rod is guiding the red light from left edge towards the right one. This test has been performed in several rods with different lengths, obtaining similar results and showing a good light guiding behavior. More interesting is to study the optical resonances overlapped on the photoluminescence spectra of the rods. To perform this experiment, a 325 nm laser is used to locally excite the PL of the rod, using a 40× objective. The PL spectra of a rod is shown in Figure 8(b), and the most intense optical resonances are indicated with arrows.

A closer inspection of the structure shows that is composed of two rods with rectangular cross section (Figure 8(b)) inset). The rod that is on the top has a section of 2.0 ×1.6 μm$^2$. The origin of the optical resonances can be related to Fabry-Pèrot modes (with the light confined between two parallel facets of the rod) or to whispering gallery modes (having total internal reflection on the four facets of the rod). The optical path that better fits with the position of the obtained optical resonances is around 5.0 μm, which corresponds to the optical path of whispering gallery modes (WGM, see scheme in Figure 9(a)).

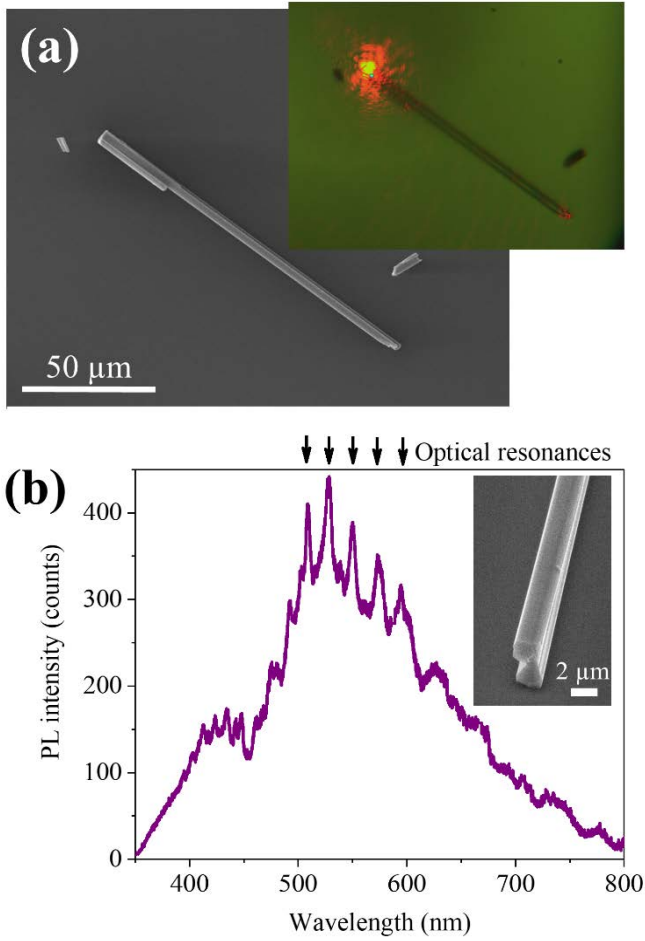

**Figure 8.** (a) 633 nm light guiding of a niobium oxide rod. (b) PL spectra recorded on a rod, with the peaks corresponding to optical resonances indicated.

The equations for the wavelength position of the WGM, depending on the polarization of the modes, in the rectangular cavity shown in Figure 10(a) are [38, 39]:

$$\lambda_{TE} = \frac{n \cdot L_{op}}{(N-2) + \frac{2}{\pi}arctan\left(n^2\sqrt{\frac{sin^2\alpha_1 - 1/n^2}{cos^2\alpha_1}}\right) + \frac{2}{\pi}arctan\left(n^2\sqrt{\frac{sin^2\alpha_2 - 1/n^2}{cos^2\alpha_2}}\right)} \quad (1)$$

$$\lambda_{TM} = \frac{n \cdot L_{op}}{N + \frac{2}{\pi}arctan\left(\sqrt{\frac{sin^2\alpha_1 - 1/n^2}{cos^2\alpha_1}}\right) + \frac{2}{\pi}arctan\left(\sqrt{\frac{sin^2\alpha_2 - 1/n^2}{cos^2\alpha_2}}\right)} \quad (2)$$

It has been reported previously that the TM modes have lower losses in the total internal reflections [40-42], so the most intense peaks are associated to this type of modes. In the spectra of Figure 8(b) some resonant peaks with lower intensity are visible between the TM modes, which could be associated with the TE modes.

Taking into account the variation of the refractive index with the wavelength through the Cauchy formula:

$$n(\lambda) = A + \frac{B}{\lambda^2} \quad (3)$$

An optical path of 5.12 μm (obtained from the size of the rod) and total internal reflection angles of α1=51.4º and α2=38.6º, the position of the resonant peaks can be fitted to equation (2). The results of the fitting are shown in Figure 9(b). The Cauchy values obtained from the fitting are A = 2.182 ± 0.007 and B = (53 ±2) ×10³ nm². These values give a refractive index at 500 nm of 2.39, very close to the values reported for niobium oxide in previous works [43-45], and it is an evidence of the high refractive index that have the obtained rods.

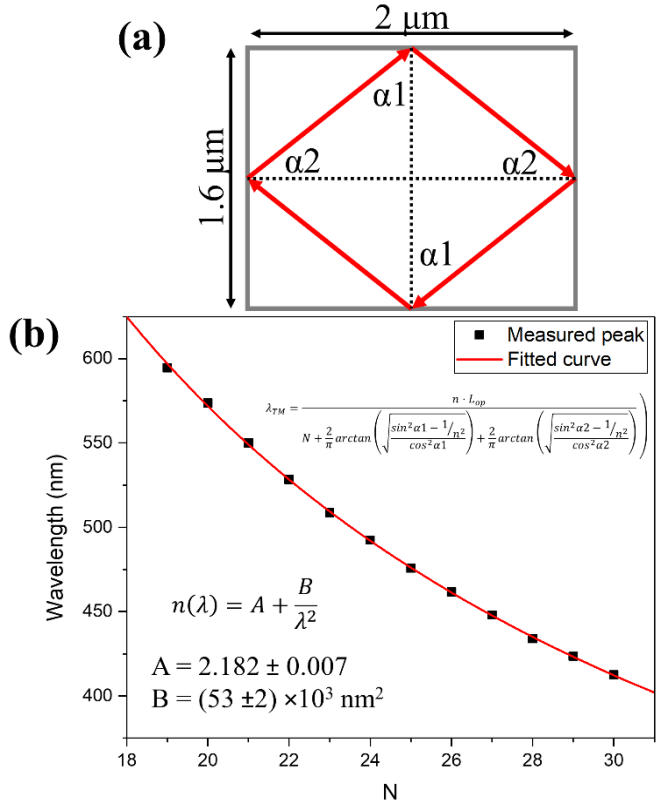

**Figure 9.** (a) Whispering gallery mode optical path in a cavity with rectangular cross-section. (b) Fit of the measured resonant peak to equations (2) and (3).

**Conclusions**

Using Tailings from the Penouta Sn-Ta-Nb Deposit as precursor, we have been able to obtain good quality microstructures of $Nb_{22}O_{54}$. A characterization of the crystal structure of the recovered material have been performed, showing that the main crystal phase present is TT-$Nb_2O_5$, from which many other phases of niobium oxide can be obtained by thermal treatments. As a proof of the possibilities that this material offers, we have grown microrods by thermal evaporation of the recovered material in an argon atmosphere. We have shown that the obtained rods are made of a non-stoichiometric phase of niobium oxide, $Nb_{22}O_{54}$, and have good crystal quality and purity. The analysis of the optical properties has shown that the refractive index of the microrods is high (and also the dielectric constant at optical frequencies), with a value of n = 2.39 at 500 nm, estimated from the position of the whispering gallery modes of a rectangular optical cavity. These results constitute a proof of concept that the recovered material from mine tailings can be used to grow non-stoichiometric niobium oxide micro and nanostructures that can be used in many other applications, from optical coatings to batteries and supercapacitors.

**Acknowledgements.** The authors are grateful to the Spanish Ministry of Science, Innovation and Universities for support via the projects ESTANNIO (RTC-2017-6629-5) and MINECO/FEDER-MAT2015-65274-R. This project has received funding from the European Union's Horizon 2020 research and innovation program under grant agreement No 776851 (carEService). B. Sotillo acknowledges financial support from Comunidad de Madrid (Ayudas del Programa de Atracción de Talento (2017-T2/IND-5465). The authors would like to thank the UCM CAI of X-Ray diffraction for performing the XRD measurements.

**Author Contributions.** B.S.: conceptualization, data curation, formal analysis, investigation, methodology, validation, writing-original draft, writing-review editing. F.A.L: conceptualization, funding acquisition, project administration, resources, writing-review editing. L.A.: investigation, methodology, writing-review editing. P.F.: conceptualization, formal analysis, funding acquisition, project administration, resources, writing-review editing. All authors have given approval to the final version of the manuscript.